\documentstyle[sprocl,psfig,epsfig]{article}

\def\be{\begin{equation}}
\def\ee{\end{equation}}
\def\bea{\begin{eqnarray}}
\def\eea{\end{eqnarray}}

\begin{document}

\title{TIME-DEPENDENT TRANSPORT IN INTERACTING MESOSCOPIC SYSTEMS}

\author{A. P. Jauho}

\address{Mikroelektronik Centret\\ Technical University of
Denmark, {\O}rsteds Plads,
Bldg. 345east,\\ DK-2800 Kgs. Lyngby, Denmark\\E-mail: antti@mic.dtu.dk}


\maketitle\abstracts{ We review recent applications of the
nonequilibrium Green function technique to time-dependent
transport in mesoscopic semiconductor systems.}

\section{Introduction}
\begin{figure}[t]
\psfig{figure=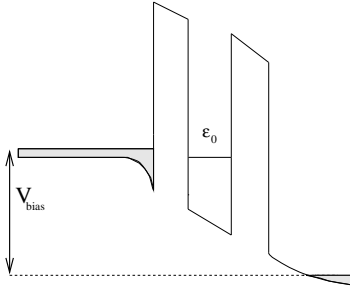,height=1.5in}
\caption{Double-barrier semiconductor heterostructure
biased close to resonance, where charge carriers emerging
from the source contact are matched to the energy of the
quasibound state $\varepsilon_0$ in the quantum well.  Occupied
contact states are shown as hatched, and the band bending
is due to charge accumulation or depletion.  \label{fig:restun}}
\end{figure}

The study of mesoscopic phenomena 
is one of the most active
areas of today's solid state physics.  One can observe
signatures of mesoscopics in a large number of different
physical systems, and a comprehensive review would not
be appropriate in the present context.
The present volume focuses on Kadanoff--Baym--Keldysh Green functions,
and consequently we restrict ourselves to certain examples
which have been studied with the
help of these techniques.
The generic system we have in mind is a semiconductor
heterostructure 
where charge carriers are introduced either
by modulation doping,
or they flow in and out of the system
through metallic or superconducting contacts.
Transport
physics in these systems can roughly be divided into two
categories: perpendicular transport
and parallel transport,
according to whether the charge carriers' motion is perpendicular
or parallel to the layers that form the heterostructures.
A representative example of perpendicular transport is
the resonant-tunneling diode (RTD),
which consists of alternate
layers of semiconductor materials with different band gaps;
a schematic conduction band diagram is shown in Fig.\ref{fig:restun}.
Charge carriers entering from left may, at a certain bias
voltage, be tuned to the quasibound state 
in the quantum
well, which results in a large enhancement of the transmitted
current.  At off-resonance conditions only a small current
can flow, because transmission through the classically
inaccessible regions is exponentially suppressed.  
This leads to a nonmonotic current-voltage
characteristic, 
and a number of device applications have
been proposed, whose operating principles are based on 
this property. 

\begin{figure}[t]
\psfig{figure=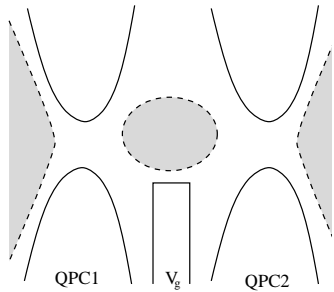,height=1.5in}
\caption{Coulomb island, which consists of two
tunable quantum point contacts
QPC1 and QPC2, and a side gate which
allows one to vary the chemical potential, and hence the
charge density in the central region.  The two-dimensional
electron gas underlying the gate structure is depleted outside
the hatched regions.\label{fig:qpc}}
\end{figure}

In the case of parallel transport much attention has been
devoted to quantum point contacts (QPC),
see Fig.\ref{fig:qpc} for
a typical experimental configuration, 
and other structures
based on similar ideas.  Here the key ingredient is metallic
gates that are deposited on the heterostructure; by adjusting
the gate potentials it is possible to deplete the underlying
two-dimensional electron gas,
and thus introduce spatial modulations
of the two-dimensional charge density.  Quantum point
contacts are based on a split gate geometry:
here, at
sufficiently high negative gate voltages, the effective connection
between the two unmodulated electron gases  (``source'' and  ``drain'')
is so narrow that perpendicular mode quantization becomes
observable, and the measured conductance is an integer
multiple of the quantum unit of conductance, $e^2/ h$.
In later sections we describe simple models pertaining to
structures like the one shown in Fig.\ref{fig:qpc}.

The hallmark of mesoscopic phenomena is the phase coherence of the
charge carriers, which is maintained over a significant part of
the transport process.  The interference effects resulting from this
phase coherence are reflected in a number of experimentally measurable
properties. 
Weak localization can be understood as an increased return
probability (and hence increased resistance) due to
{\it coherent} backscattering of charge carriers. 
Another example where phase coherence is central
is the Aharonov--Bohm effect,
where interference of two different transport
paths in a ring geometry results in an oscillatory
magnetoresistance.
Yet another example is universal conductance
fluctuations, 
where the conductance of a sample displays rapid
changes on the scale of $e^2/h$ (hence the ``universality")
when an external control parameter is changed.  The external
parameter could be magnetic field, or thermal cycling, and
the fluctuations reflect changes in the conducting channels
either due to different impurity configurations (thermal
cycling), or differences in the way the conduction channels
are located in the sample (magnetic field).
In these notes
we focus on an alternative way of affecting the
phase coherence: external {\it time-dependent} perturbations.
The interplay of external time dependence and phase coherence
can be phenomenologically understood as follows.  If the single-particle
energies acquire a time dependence, then the wavefunctions have
an extra phase factor, $\psi \sim \exp[-i\int^t d t'\varepsilon(t')]$.
For a uniform system such an overall phase factor is of no
consequence. However, if the external time dependence is
different in different parts of the system, and the particles
can move between these regions (without being ``dephased"
by inelastic collisions), the phase difference becomes
important. 

The interest in time-dependent mesoscopic phenomena stems from 
recent progress in several experimental 
techniques.  Time dependence
is a central ingredient in many different experiments,
of which we mention the following:
(i) {\it Single-electron pumps and turnstiles.}  
Here, time-modified gate signals move electrons one by one through
a quantum dot, leading to a current which is proportional to the
frequency of the external signal.  These structures have considerable
importance as current standards. 
The Coulombic repulsion
of the carriers in the central region
is crucial to the operational principle of these devices, and
underlines the fact that extra care must be paid to interactions when
considering time-dependent transport in mesoscopic systems.
(ii) {\it ac response and transients in resonant-tunneling devices.}
Resonant tunneling devices (RTD) have a number of applications as
high-frequency amplifiers or detectors.  For the device engineer,
a natural approach would be to model these circuit elements with
resistors, capacitances, and inductors.  The question then arises
as to what, if any, are the appropriate ``quantum" capacitances and
inductances
one should ascribe to these devices.  Answering this question requires
the use of time-dependent quantum-transport theory.

A central issue will be the treatment of interactions
in the mesoscopic region, and, as we shall see,
nonequilibrium Green function techniques
are well suited for this purpose.
First analyses of
tunneling problems
in nonequilibrium systems were performed already
in the 70s by Caroli and
co-workers 
\cite{Caroli1,Caroli2,Caroli3,Combescot}.  
In later years the steady state situation
has been addressed by a large number of papers,
however the literature on time-dependent nonequilibrium transport
treated with Green functions
is much more restricted than in the stationary case (see the recent
text-book by Haug and Jauho \cite{Haug}
for extensive references for works
published before 1996; this review will concentrate on some of the
more recent advances.  Additional useful and related information
can also be found in the monograph by Datta\cite{Datta}.). 

Among the central results obtained 
with the nonequilibrium Green function techniques
is that under certain conditions (to be discussed below)
a Landauer-type conductance formula 
\cite{Landauer1,Landauer2}  can be
derived.  
The Landauer formula
relates the conductance $g$ of a mesoscopic sample
[which is connected via ``ideal" leads to two (or more) reservoirs]
to its transmission properties, $g = (e^2/h) T$, where $T$ is
the quantum mechanical transmission coefficient
of the sample.
Conductance formulae  have played an important role in
the analysis of many mesoscopic transport phenomena, and
it is therefore of interest to investigate whether interactions and/or
time-dependence can be treated in a similar fashion.  This study
forms the core of the present review.

\section{Nonequilibrium formulation of tunneling physics}
The total current measured in an external ammeter can be split
into two contributions: the current flowing into the mesoscopic
region, and the current flowing in and out of the accumulation or
depletion regions in the leads, i.e., the displacement or 
capacitive current.  
We begin by  deriving an expression for the tunneling current, and comment
on models for the displacement current later.  One should also note
that the displacement current does not contribute to the time averaged
current, which is often the experimentally relevant quantity.  We shall
also assume that the electric fields in the leads are effectively screened
out, so that the voltage drop occurs across the mesoscopic region.
This assumption sets an upper bound for the 
external driving frequency, which should not
exceed the plasma frequency.  Estimates for the cut-off frequency are
in the range of 10 GHz -- 1 THz, depending on the device geometry and
dimensionality \cite{JWM94}, which are sufficiently high for most present
applications.

We recall that the basic difference between 
construction of
equilibrium
and nonequilibrium perturbation schemes is that in nonequilibrium
one cannot assume that the system returns to its ground
state (or a thermodynamic equilibrium state at finite
temperatures) as $t\to +\infty$.  Irreversible
effects break the symmetry between $t=-\infty$ and
$t=+\infty$, and this symmetry is heavily
exploited in the derivation of the equilibrium perturbation
expansion \cite{Mahan,FW,Enz}.  In nonequilibrium situations one can circumvent
this problem by allowing the system to evolve from
$-\infty$ to the moment of interest (for definiteness,
let us call this instant $t_0$), and then continues the
time evolvement from $t=t_0$ back to 
$t=-\infty$. 
The advantage of this 
procedure is that all expectation values are defined
with respect to a well-defined state, i.e., the state
in which the system was prepared in the remote past.  The
price  is that one must treat the  
two time branches
on an equal footing.

In the context of tunneling problems 
the nonequilibrium formalism
works as follows.  In the remote past the contacts (i.e., the
left and right lead) and the central region are decoupled,
and each region is in thermal equilibrium.
The equilibrium distribution functions for the three regions
are characterized by their respective chemical potentials;
these do not have to coincide nor are the differences between
the chemical potentials necessarily small.  The couplings between the
different regions 
are then established and treated as perturbations 
via  the standard techniques of perturbation
theory, albeit on the two-branch time contour.
It is important
to notice that the couplings do not have to be small,
e.g., with respect level to spacings or $k_{\rm{B}} T$, and typically
must be treated to all orders.

\section{Current formulas}\label{sect:current}

We can now present the mathematical formulation of the problem. 
A detailed presentation can be found in a recent text-book \cite{Haug}, and
here we present just some of the central ideas. The
contacts are assumed to be noninteracting, and the single-particle
energy in lead $\alpha$ is given by
\begin{equation}
\varepsilon_{k,\alpha}(t)=\epsilon_{k,\alpha} + \Delta_\alpha(t)\;,
\end{equation}
where $\Delta_\alpha(t)$ is the external time modulation.
The leads are connected to the central (or, mesoscopic) region
via a hopping term with matrix element $V_{k\alpha;n}(t)$, where $n$
labels the eigenstates of the central region.  Collecting the various
terms results in the Hamiltonian $H=H_L+H_R+H_T+H_{\rm cen}$, or,
explicitly:
\begin{equation}
H = \sum_{k,\alpha} \epsilon_{k,\alpha}(t) c^\dagger_{k,\alpha} c_{k,\alpha}
+ \sum_{k,\alpha;n}
\left[V_{k\alpha;n}(t) c^\dagger_{k,\alpha} d_n + {\rm h.c.}\right]
+ H_{\rm cen}\left[\{d_n\},\{d^\dagger_n\},t\right]\;,
\end{equation}
where the central part Hamiltonian must be chosen according to the
system under consideration.  The operators
$\{d_n\},\{d^\dagger_n\}$ refer to a complete set of
single-particle states of the central region.
The derivation of the
basic formula for the time-dependent
current does not require an explicit form for  $H_{\rm cen}$; the
actual evaluation of the formula of course requires this information.  We write
$H_{\rm cen}=\sum_n \epsilon_n(t) d^\dagger_n d_n +
H_{\rm int}$, where $H_{\rm int}$
could be electron-phonon interaction, or an Anderson impurity:
\begin{eqnarray} 
H_{\rm int}^{\rm el-ph} &=& \sum_{n\sigma} 
d^{\dagger}_{n,\sigma} d_{n,\sigma} \sum _{\bf q} M_{n,\bf q}
\left[a^\dagger_{\bf q} + a_{\bf q}\right]\label{Helph}\\
H_{\rm int}^A &=& U \sum_n
d_{n,\uparrow}^\dagger d_{n,\uparrow} d_{n,\downarrow}^\dagger d_{n,\downarrow}
\label{HAnd}\;.
\end{eqnarray}
According to the basic ideas of the tunneling approach presented
above, the occupations
of the leads are determined by equilibrium distribution functions.
Thus the
Green functions for the contacts are known explicitly:
\begin{eqnarray}
g_{k\alpha}^<(t,t') &\equiv&
i\langle{\bf{c}}_{k\alpha}^{\dagger}(t'){\bf{c}}_{k\alpha}(t)\rangle
\nonumber\\
&=&
i f(\varepsilon^0_{k\alpha})
\exp\big [ -i \int_{t'}^t d t_1\varepsilon_{k\alpha}(t_1)\big ]\;,
\label{gcontactless} \\
g_{k\alpha}^{r,a}(t,t') &\equiv& \mp i \theta(\pm t \mp t')
\langle\lbrace{\bf{c}}_{k\alpha}(t),
{\bf{c}}_{k\alpha}^{\dagger}(t')\rbrace\rangle 
\nonumber\\
&=&
\mp i\theta(\pm t \mp t')
\exp\big [ -i \int_{t'}^t dt_1\varepsilon_{k\alpha}(t_1)\big ] \;.
\label{gcontactr}
\end{eqnarray}
We start the derivation by considering
the current leaving the, e.g., left contact, and entering
the central region:
\begin{equation}
J_L(t) = \langle I_L(t) \rangle
= \langle (-e) {\dot N}_L (t) \rangle = 
-ie \langle \left[ H,N_L \right]\rangle \;.
\end{equation}
The commutator $[H,N_L]$ is readily evaluated, and one finds
\begin{equation}
J_L(t) = {2e\over\hbar}
{\rm Re} \Big\{
\sum_{k,\alpha,n} V_{k\alpha,n}(t) G^<_{n,k\alpha}(t,t)\Big \}\;,
\end{equation}
which involves the time-diagonal part of the correlation function
\begin{equation}
G^<_{n,k\alpha}(t,t') = i \langle c^\dagger_{k\alpha}(t') d_n (t) \rangle\;.
\end{equation}
The next step consist of writing down the equation-of-motion for
the time-ordered function
$G^t_{n,k\alpha}(t,t')$, and a subsequent
analytic continuation with the Langreth rules \cite{LangrethRules} leads to
\begin{equation}
G^<_{n,k\alpha}(t,t') = \sum_m \int dt_1 V^*_{k\alpha,m }
\Big [ G^r_{nm}(t,t_1) g^<_{k\alpha}(t_1,t')
+ G^<_{nm}(t,t_1) g^a_{k\alpha}(t_1,t') \Big ].
\end{equation}
Substituting the expressions for the contact Green functions, 
Eqs.(\ref{gcontactless})--(\ref{gcontactr}) finally yields
\begin{eqnarray}
J_L(t) &=& - {2e\over \hbar} \int_{-\infty}^t dt_1 \int {d\epsilon\over 2\pi}
{\rm ImTr} \Big \{ e^{-i\epsilon(t_1-t)} {\bf \Gamma}^L(\epsilon,t_1,t)
\nonumber\\
&\quad&\quad\times
\left[ {\bf G}^<(t,t_1) + f^0_L(\epsilon) {\bf G}^r(t,t_1)\right] \Big \}\;.
\label{jfinal}
\end{eqnarray}
Here the Green functions ${\bf G}^{<,r}$ are {\it matrices} in the
indices $(m,n)$, and the linewidth functions ${\bf \Gamma}$ are defined as
\begin{equation}
\left[ {\bf \Gamma}^L(\epsilon,t_1,t)\right]_{mn}
= 2\pi\sum_{\alpha\in L}
\rho_\alpha(\epsilon) V_{\alpha,n}(\epsilon,t) V^*_{\alpha,m}(\epsilon,t_1)
\exp\big[-i \int_t^{t_1} dt_2 \Delta_\alpha (\epsilon,t_2)\big]\;,
\end{equation}
where $\rho_\alpha(\epsilon)$ is the density of states.
Equation (\ref{jfinal}) is the main formal result of this report: 
the subsequent sections
are devoted to exploring its special applications.  It is important to note
that the current formula only involves the Green function of the central
region.  However, ${\bf G}^<(t,t_1)$ must be calculated in the presence
of the coupling to the leads, which is a highly nontrivial task for an
interacting system.  Thus, Eq.(\ref{jfinal}) can be viewed as a rather formal
statement, but nevertheless it provides under many circumstances
a very convenient starting-point
for further calculations.
\section{Special cases}
\subsection{Stationary limit}
In the stationary limit Eq.(\ref{jfinal}) can be further simplified and
we get the result first reported in Ref.\cite{MW}:
\begin{eqnarray}
J &=& {ie\over 2\hbar} \int {d\epsilon\over 2\pi} {\rm Tr}\Big \{
\left[ {\bf \Gamma}^L(\epsilon) - {\bf \Gamma}^R(\epsilon)\right]
{\bf G}^<(\epsilon)\nonumber\\
&\quad&\quad +
\left[ f_L^0(\epsilon){\bf \Gamma}^L(\epsilon) - f_R^0(\epsilon)
{\bf \Gamma}^R(\epsilon)\right]
\left[ {\bf G}^r(\epsilon) - {\bf G}^a(\epsilon)\right]\Big\}\\
&=&  {i e\over\hbar}\int{d\varepsilon\over 2\pi}
\left[f_L(\varepsilon)-f_R(\varepsilon)\right] 
{\cal T}(\varepsilon)\;,
\label{jprop}
\end{eqnarray}
where
\begin{equation}
{\cal T}(\varepsilon)  = 
{\rm {Tr}} 
\left\lbrace {{\bf \Gamma}^L(\varepsilon){\bf \Gamma}^R(\varepsilon)\over 
{\bf \Gamma}^L(\varepsilon)+{\bf\Gamma}^R(\varepsilon)}
\bigl[{\bf  G}^r(\varepsilon)-{\bf  G}^a(\varepsilon)\bigr ]\right\rbrace
\;.
\label{calT}
\end{equation}
Equation (\ref{jprop}) holds for the special case when the couplings
between the left and right leads are proportional (a constant coupling,
occurring in the wide-band limit is a frequently encountered special case).
The ratio in the curly brackets in Eq.(\ref{calT}) is well-defined because 
the ${\bf {\Gamma}}$-matrices  are, by assumption, proportional.
The difference between the retarded and advanced Green functions
is essentially the density of states. Despite the apparent
similarity 
of Eq.(\ref{jprop}) to the Landauer formula,
it is important to bear in
mind that, in general, there is no immediate connection between
the quantity 
$\cal T(\varepsilon)$
and
the transmission coefficient $T(\varepsilon)$. 
In particular, when inelastic scattering
is present, there is no such  connection.
These expressions for the stationary current have a wide range of
applicability
for calculations of current-voltage relations in mesoscopic structures.
Recent applications include nonlinear current-voltage calculations
\cite{Wang}, 
transport in carbon nanotubes \cite{Nardelli,Anantram},
Kondo physics \cite{Yeyati,Takagi,Sakai,Thimm,Hettler},
Andreev scattering in semiconductor--superconductor hybrid systems
\cite{Raimondi}
(we return to these two topics below),
tunneling through magnetic barriers
\cite{Heide},
delocalization of excitons in disordered systems
\cite{Berkovits},
role of correlations in transport through quantum dots or
artificial molecules
\cite{Craco,Akera,Georges,Ramirez,Jonault,Stafford,Favand},
physics of nanowires
\cite{Cuevas,Mads},
analysis of STM (Scanning Tunneling Microscope) experiments
\cite{Briggs,STM},
and many others.

\subsection{Noninteracting electrons}\label{subsec:free}
\begin{figure}[t]
\psfig{figure=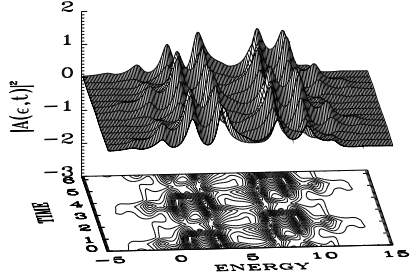,height=1.5in}
\caption{$|A(\epsilon,t)|^2$ as a function of time for harmonic modulation
for a symmetric structure, $\Gamma_L=\Gamma_R=\Gamma/2$.
The unit for the time-axis is $\hbar/\Gamma$, and all energies
are measured in units of $\Gamma$, with the values
$\mu_L=10$, $\mu_R=0$, $\epsilon_0=5$, $\Delta=5$, $\Delta_L=10$,
and $\Delta_R=5$.  The modulation frequency is
$\omega=2\Gamma/\hbar$. \label{fig:AbsA}}
\end{figure}

\begin{figure}[t]
\psfig{figure=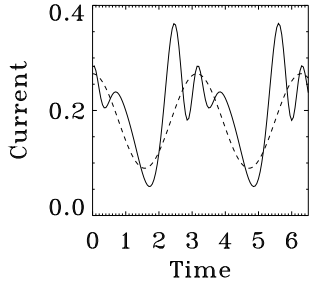,height=1.5in}
\caption{The time-dependent current $J(t)$ for harmonic modulation.
The dc bias is defined via $\mu_L=10$ and $\mu_R=0$, respectively.  
The dotted line
shows (not drawn to scale) the time dependence of the drive signal.
The temperature is $k_BT=0.1\Gamma$.\label{fig:harmcur}}
\end{figure}

Now $H_{\rm cen}=\epsilon_0 (t)d^\dagger d = 
[\epsilon_0 + \Delta_0(t)]d^\dagger d$ (we
examine only the single-level case; a generalization to many
levels and/or spin is straightforward, all the results are formally
the same but the Green functions, self-energies etc. must
be interpreted as matrices), and we can give the following
explicit results for the occupation $N(t)=-iG^<(t,t)$, and
the current $J_{L/R}(t)$ \cite{JWM94}:
\begin{eqnarray}
N(t) &=& \sum_{L/R} \int {d\epsilon\over 2\pi} f^0_{L/R}(\epsilon)
|A_{L/R}(\epsilon,t)|^2\\
J_{L/R}(t) &=& - {e\over \hbar} \Gamma_{L/R} \Big [
N(t) + {1\over \pi} \int d\epsilon f^0_{L/R}(\epsilon)
{\rm Im}\left\{ A_{L/R}(\epsilon,t)\right\} \Big ]\;,
\end{eqnarray}
where
\begin{eqnarray}
A_{L/R}(\epsilon,t) &=& \int dt_1 e^{i\epsilon(t-t_1)}
e^{-i\int_t^{t_1}dt_2 \Delta_{L/R}(t_2)}
G^r(t,t_1)\nonumber\\
&=&e^{-i{\Delta_0\over\gamma}\sin(\gamma t)}
\sum_{k=-\infty}^\infty 
{J_k\left[(\Delta_0-\Delta_{L/R})/ \gamma\right]
e^{ik\gamma t}\over \epsilon-\epsilon_0 -k\gamma + i \Gamma/2}
\;,
\end{eqnarray}
where $J_k$ is the k:th order Bessel function, and 
we assumed a harmonic time-variation:
$\Delta_0(t)= \cos\left(\gamma t\right)$.  Figures \ref{fig:AbsA} and
\ref{fig:harmcur} give a numerical example.  We draw attention 
to the maxima in the plot for $|A|^2$; these are related to
photonic side-bands occurring at $\epsilon=\epsilon_0 \pm k\omega$
\cite{BL82}.  More work along similar lines can be found, e.g., in
Ref.\cite{AIP}
Bearing in mind the complex structure seen in Fig. \ref{fig:AbsA} it is
not surprising that the current in Fig. \ref{fig:harmcur} displays
a non-adiabatic time-dependence.
The basic physical mechanism underlying the secondary maxima and
minima in the current is the line-up of a photon-assisted
resonant tunneling peak with the contact chemical potentials.

\subsection{Average current}\label{subsec:avecur}
For constant couplings to leads a very simple result can be
obtained for the average current:
\begin{equation}
\langle J_L(t) \rangle = {\Gamma_L \Gamma_R\over \Gamma_L + \Gamma_R} 
{\rm Tr}\left\{ - {2e\over\hbar} \int {d\epsilon\over 2\pi}
\langle {\rm Im} {\bf A}(\epsilon,t)\rangle
\left[ f^0_L(\epsilon) - f^0_R(\epsilon) \right] \right\}\;.
\end{equation}
In the noninteracting single-level case $\langle {\rm Im} A(\epsilon,t)\rangle$
is given by
\begin{equation}
\langle {\rm Im} A_{L/R}(\epsilon,t)\rangle = - {\Gamma\over 2}
\sum_{k=-\infty}^{\infty}
{J_k^2\left[(\Delta_0-\Delta_{L/R})/\omega\right] \over
(\epsilon-\epsilon_0-k\omega)^2 + (\Gamma/2)^2} \;.
\nonumber
\end{equation}
We shall use these expressions below.

\subsection{Resonant tunneling with phonons}
Resonant tunneling diodes (see Fig. \ref{fig:restun})
are important both for technical applications,
such as oscillators or infra-red detectors, but they are equally important
conceptually as a very clear-cut example of a system exhibiting
interacting, far-from equilibrium
transport phenomena.  In this subsection we consider electron-phonon
interactions.  As explained in the Introduction, the IV-characteristic
is dominated by a maximum when the quasibound state is aligned
with the energies of the incoming electrons.
Experimentally, however, it was found \cite{Goldman} that the
current-voltage characteristic could also display a secondary maximum,
i.e. a satellite of the main feature.  The reason is interaction
with phonons: an electron, which
approaches the double-barrier structure with a nonresonant
energy, can be tuned in the energy of the quasibound state
in the quantum well by emitting (or absorbing) an optical
phonon, and thus become resonant with an enhanced transmission
probability, and increased current.  The central-region
Hamiltonian (\ref{Helph}) is a mathematical formulation for
this physical picture. 

For simplicity, we consider
only energy-in\-de\-pen\-dent
level-widths $\Gamma_L$ and $\Gamma_R$,   
when the current (\ref{jprop}) becomes 
\begin{equation}
J = {e\over\hbar}{\Gamma^L\Gamma^R \over \Gamma^L+\Gamma^R}
\int {d\varepsilon\over 2\pi} [f_L(\varepsilon)-f_R(\varepsilon)]
\int_{-\infty}^{\infty}dt e^{i\varepsilon t}A(t)\;,
\label{Jelph}
\end{equation}
where $A(t)=i[G^r(t)-G^a(t)]$ is the interacting
spectral function. 
In general, an exact evaluation of $A(t)$
is not possible, however, if one neglects the coupling to
the contacts (or treats the coupling phenomenologically by introducing
an exponential decay in $A(t)$), $G^r(t)$ 
[and hence $A(t)$] can be calculated
exactly \cite{Mahancalc}.  A very convenient way to calculate $A(t)$ is
to use the linked-cluster theorem \cite{Mahan}.  
We write the electron-phonon interaction
in the interaction picture (for simplicity we again consider just one level),
\begin{equation}
H_{\rm int}^{\rm el-ph}(t) = d^{\dagger} d \sum _{\bf q} M_{\bf q}
\left[a^\dagger_{\bf q}e^{-i\omega_q t} + a_{\bf q}e^{i\omega_q t}
\right]\;.\label{Helphint}
\end{equation}
$H_{\rm int}^{\rm el-ph}(t)$ can then be viewed as an effective one-electron
Hamiltonian in a time-dependent Schr{\"o}dinger equation, the solution of
which gives $A(t)$ (the averaging is now over the phonon subsystem):
\begin{eqnarray}
A(t)&=&e^{-i\epsilon_0 t}
\langle T\{ \exp[-i\int_0^t dt_1 H_{\rm int}^{\rm el-ph}(t_1)]
\}\rangle
\nonumber\\
&=&e^{-i\epsilon_0 t}\exp[-{1\over 2} \int_0^t dt_1 \int_0^t dt_2 \langle T\{
H_{\rm int}^{\rm el-ph}(t_1)H_{\rm int}^{\rm el-ph}(t_2)\}\rangle]\;,
\end{eqnarray}
where $T$ is the time-ordering operator, and
we used the linked-cluster theorem on the second line.
The time-ordered product is nothing but the free phonon Green function:
\begin{equation}
i\langle T\{
H_{\rm int}^{\rm el-ph}(t_1)H_{\rm int}^{\rm el-ph}(t_2)\}\rangle
= (N+1)e^{-i\omega_q|t_1-t_2|} + N e^{i\omega_q |t_1-t_2|}\;,
\end{equation}
where $N=1/[\exp(\hbar\omega_q\beta)-1]$.  The time-integrals are easily
worked out with the result
\begin{equation}
A(t)  =  \exp [-i t(\varepsilon_0-\Delta)-\Phi(t)-\Gamma |t|/2]\;,
\end{equation}
where
\begin{eqnarray}
\Delta  &=& \sum_{{\bf  q}} {M_{{\bf  q}}^2 \over \omega_{{\bf  q}}}\;,\\
\Phi(t)  &=& \sum_{{\bf  q}}{M_{{\bf  q}}^2 
\over \omega_{{\bf  q}}^2}[N_{{\bf  q}}
(1-e^{i\omega_{\bf  q}t})+(N_{\bf  q}+1)(1-e^{-i\omega_{\bf  q}t})]\;.
\end{eqnarray}
When substituted in the expression for current, one recovers the result
of Wingreen et al.\cite{WJW}, which originally was derived by analyzing
the much more complex two-particle Green function 
\begin{equation}
G(\tau,s,t)=\theta(s)\theta(t)
\langle d(\tau-s)d^{\dagger}(\tau)d(t)d^{\dagger}(0)\rangle. 
\end{equation} 
The advantage of the method presented here is that
one only needs the {\it single}-particle Green function to use
the interacting current formula (\ref{jprop}).
Other systematic approaches to the single-particle Green
function can therefore be directly applied to the current
(e.g., perturbation theory in the tunneling Hamiltonian;
Anda and Flores\cite{Anda}, Hyldgaard et al.\cite{Hyld}).
The model studied in this section is quite flexible and it has
been used to describe many other interaction mechanisms in double-barrier
systems, such as light assisted magnetotunneling \cite{IP}, or
plasmon assisted tunneling \cite{Zhang}.

\subsection{Multiterminal generalization and displacement current}

We now return to the issues of current partition 
and   displacement current in a multiprobe sample, mentioned
in the Introduction.  
The dc situation has been exhaustively analyzed, starting
from the seminal work of B{\"u}ttiker et al \cite{BILS}.
The time-dependent case is more complicated: not only has the
tunneling current to be considered, but also the displacement
contributions must be accounted for.  These issues are essential
for two reasons: i) Current conservation and ii) Gauge invariance.
(Gauge invariance means that a uniform shift of all voltages should
not affect the final results.)
A linear, low frequency theory was developed by B{\"u}ttiker and co-workers
\cite{Betal}, based on an generalization of the scattering matrix
formulation of transport theory.
Nonequilibrium Green functions allow, at least in principle, the
analysis of high frequencies and far-from equilibrium situations.
Here we will briefly review recent progress within this formulation.
Stafford discussed the current partition in dc but under nonlinear
conditions \cite{Stafford96}, and recent work by 
Anantram and Datta \cite{Ana95} and Wang et al.\cite{Wang99}
have led to an ac generalization.
Concerning the tunneling contribution, it is a trivial matter to
generalize from the two-probe geometry to a multiprobe system:
one merely replaces the indices $L,R$ by $\alpha$, where $\alpha$
labels the probes, and considers
the current $J_\alpha(t)$ (or $J_\alpha(\omega)$, whichever is
more convenient).
The dynamic conductance $G_{\alpha\beta}$
due to the tunneling current is defined
as  
\begin{equation}
J_\alpha(\omega)=\sum_\beta G_{\alpha\beta}(\omega)V_\beta(\omega).
\end{equation}
In the time-dependent case the tunneling currents do not add up
to zero, due to charge accumulation/depletion.  The total current,
however, is conserved:
\begin{equation}
\sum_\alpha J_\alpha(\omega) = i\omega Q(\omega)\;,
\label{sum}
\end{equation}
where
\begin{equation}
Q(\omega) = -\sum_\beta iq \int {dE\over 2\pi}
{\rm Tr}[g^<_\beta(E+\omega,E)]V_\beta
\end{equation}
is the accumulated charge in the scattering region; 
$g^<_\beta (E,E')$ is the double Fourier transform of
the {\it small signal} component of the full Green function
$G^<$, See Refs.[\cite{Ana95,Wang99}].  The total current in probe $\alpha$
is $J_\alpha^{\rm tot}=J_\alpha+J_\alpha^d$, where 
$\sum_\alpha J_\alpha^d=J^d=dQ/dt$, and current conservation means
$\sum_\alpha J_\alpha^{\rm tot}=0$.  Additional information is
required to partition $J^d$, because only the sum of
the various displacement currents is
known via Eq.(\ref{sum}).  In a model where coupling constants
between the central region and the contacts are independent of
energy, one can readily do this partitioning: $J^d_\alpha=
(\Gamma_\alpha/\sum_\beta\Gamma_\beta)J^d$.
In a more elaborate model Wang et al.\cite{Wang99} have
outlined a procedure how this partitioning can be carried out;
the analysis is based on requirements of charge conservation,
$\sum_\alpha G_{\alpha\beta}^{\rm tot}=0$, and gauge invariance,
$\sum_\beta G_{\alpha\beta}^{\rm tot}=0$.  The end result for
the dynamical conductance is
\begin{equation}
G_{\alpha\beta}^{\rm tot}(\omega)=G_{\alpha\beta}
- G^d_\beta{\sum_\gamma G_{\alpha\gamma}\over G^d_\gamma}\;,
\label{dyncond}
\end{equation}
where 
\begin{equation}
G_\beta^d = -q\omega \int{dE\over 2\pi} {\rm Tr}
[g^<_\beta(E+\omega,E)]\;.
\end{equation}
The result (\ref{dyncond}) formally agrees with the scattering
matrix results of B{\"u}ttiker et al. \cite{BILS}, but now the various terms
are expressed in terms of nonequilibrium Green functions, and 
hence powerful techniques for evaluating them are available.
Clearly the final word is not said in this rapidly developing
subfield.

\subsection{Phase-measurement of a quantum dot}

\begin{figure}[t]
\psfig{figure=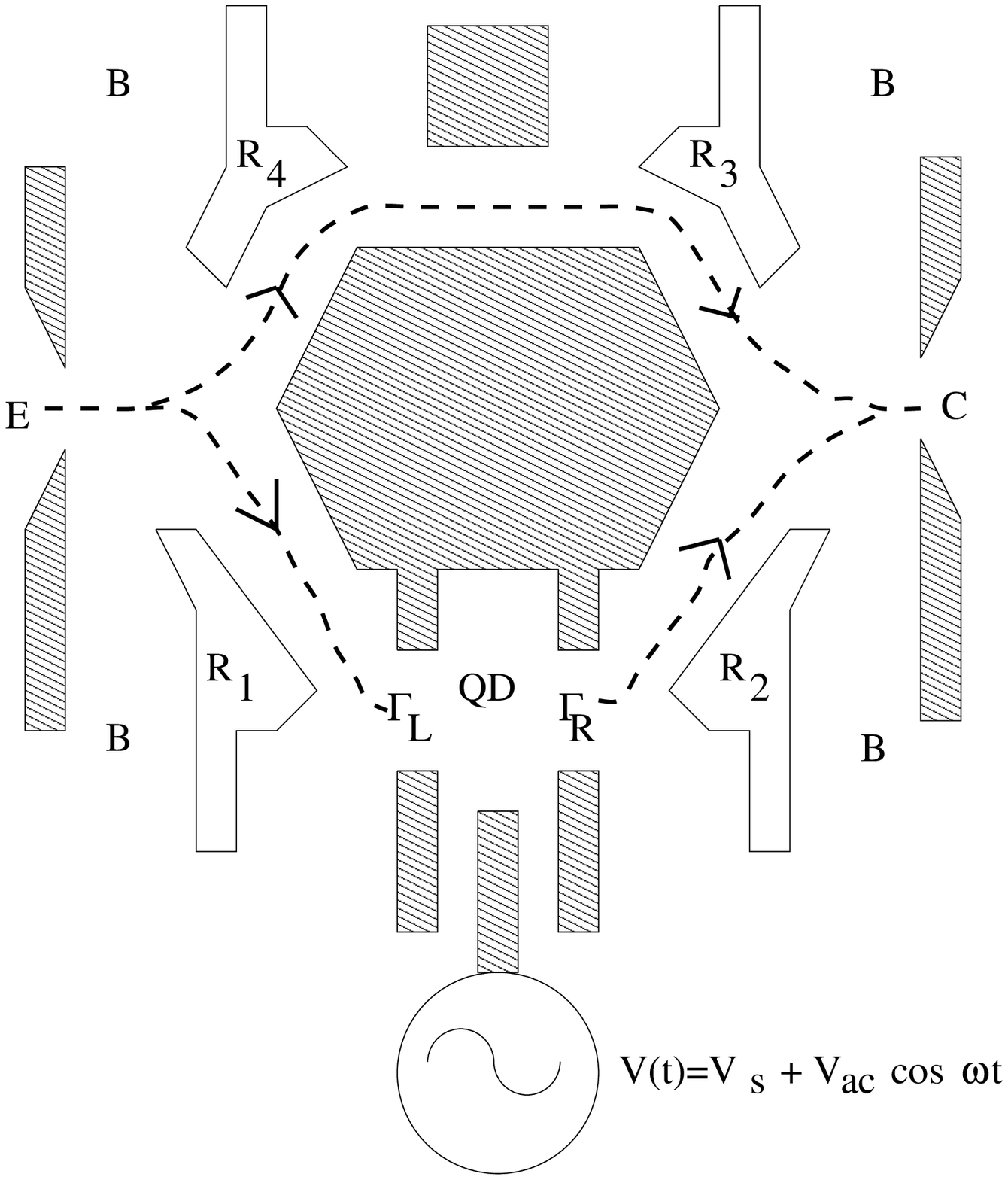,height=1.5in}
\caption{Schematic lay-out of the proposed 
multi-terminal double-slit interference 
experiment. The device consists of an Aharonov-Bohm ring with a
quantum dot (QD) in one arm, defined with metallic gates (shaded
areas) on the 2D electron gas. Electron paths (shown as dashed lines)
originating from the
emitter (E) interfere at the collector (C).  
A time-dependent voltage $V(t)$ is applied
to the quantum dot via a side gate. 
The reflector gates $R_{1\cdots 4}$, shown as white areas, 
direct multiply reflected
paths to the common base (B), thus preventing them
from contributing to the interference signal, in accordance
with Eq.(\ref{Tring}).}
\label{f1}
\end{figure}

Measuring the phase of a transmission coefficient
in contrast to the amplitude (which determines conductance)
has become possible only quite recently
\cite{Yacoby,Schuster}. The experimental protocol can be
summarized as follows:
A magneto-transport measurement is performed on an Aharonov--Bohm ring
with a quantum dot fabricated in one of its arms.
If the quantum dot supports coherent transport, the transmission
amplitudes through the two arms interfere. 
A magnetic field induces
 a relative phase change, $2\pi \Phi/\Phi_0$, between the
two transmission amplitudes, $t_0$ and 
$\tilde{t}_{\scriptscriptstyle{\!Q\!D}}$,
leading to an oscillatory component to the
conductance $g(B) = (e^2/h){\cal T}(B)$, with 
\begin{equation}
{\cal T}(B)=
{\cal T}^{(0)} +
2{\rm Re}\{ t^*_0 
\tilde{t}_{\scriptscriptstyle{Q\!D}} e^{2\pi i \Phi/\Phi_0}\}
+...,
\label{Tring}
\end{equation}
where $\Phi$ is the flux threading the ring, $\Phi_0=hc/e$ is the flux quantum, 
and where the dots represent higher harmonics due to multiple 
reflections.
The amplitudes 
$t_0$ and $\tilde{t}_{\scriptscriptstyle{\!Q\!D}}$ give the
{\em coherent} parts of the two sets of paths joining
the emitter and the collector; the incoherent
components lead to a structureless background signal, which
can be neglected in the forthcoming analysis.
In the experiments, an oscillatory component in
magnetoconductance of this form
was clearly observed thus demonstrating coherent transmission through
the arm with the dot \cite{Yacoby,Schuster}. 
In the experiment of Yacoby {\it et al.} \cite{Yacoby}, the Aharonov--Bohm
phase could take on only two values, $0$ and $\pi$, as a consequence
of microreversibility in a two-terminal geometry. 
The second
generation of experiments \cite{Schuster}, in a four-terminal geometry, allowed
the determination of the continuous phase shift of the transmission
amplitude 
through the dot.
The success of these experiments 
suggests  applications
to other phase-coherent transport 
processes. One particular example   which has been of considerable recent
interest
is photon-assisted tunneling. 
While photon-assisted tunneling (PAT) is intrinsically a 
coherent phenomenon, existing measurements
of PAT are insensitive to the phase of the transmitted electrons
and do not directly demonstrate coherence in the presence of
the time-dependent field. Here we give a brief account of
a recent proposal \cite{JW} for a measurement of
photon-assisted tunneling through a quantum dot 
in the mesoscopic double-slit geometry described above
(see Fig. \ref{f1}). 
The proposed experiment is a combination of the experiments of
Kouwenhoven {\it et al.} \cite{Leo,Ooster} 
where a microwave modulated side-gate voltage gave rise
to photon-assisted tunneling through a quantum dot, and the
interference experiments 
of Refs. \cite{Yacoby} and \cite{Schuster}.

We can use the results presented in Section \ref{subsec:free}
to construct an expression for the interference signal.
We focus on transport in the neighborhood of a single Coulomb
oscillation peak associated with a single nondegenerate electronic
level of the quantum dot. The effect of the ac side-gate voltage is
described entirely through the time-dependent energy of this level,
which has the familiar form
\begin{equation}
\epsilon(t)=\epsilon_0(V_{\rm s})+V_{\rm ac}\cos \omega t, 
\label{energy}
\end{equation}
where the static energy of the level $\epsilon_0$ depends on the dc side-gate
voltage $V_{\rm s}$.
 All other levels on the
dot can be neglected provided  the ac amplitude, $V_{\rm ac}$,
 and the photon energy, $\hbar \omega$, are small
compared to the level spacing on the dot.

The energy-dependence of the coherent part of the transmission amplitude
$\tilde {t}_{\scriptscriptstyle{Q\!D}}(\epsilon)$ through
the arm containing the quantum dot is determined by
the transmission amplitude $t_{\scriptscriptstyle{Q\!D}}(\epsilon)$
through the dot,
$\tilde {t}_{\scriptscriptstyle{Q\!D}}(\epsilon)\propto
t_{\scriptscriptstyle{Q\!D}}(\epsilon)$.
In the absence of an ac potential, a suitable model for 
the dot transmission amplitude 
is the Breit--Wigner form,
\begin{equation}
t_{\scriptscriptstyle{Q\!D}}(\epsilon)={{-i\sqrt{\Gamma_L \Gamma_R}} \over 
{\epsilon - \epsilon_0(V_{\rm s}) + i\Gamma/2}}\;,
\label{tqdstatic}
\end{equation}
where 
$\Gamma=\Gamma_L+\Gamma_R$
is the full width at half maximum of the resonance 
on the dot due to tunneling to the left and right leads.
 Eq. (\ref{tqdstatic}) implies 
a continuous phase accumulation  of $\pi$ 
in the transmission amplitude as the Coulomb
blockade peak is traversed. (Note that 
the Breit--Wigner form is exact for a noninteracting system with $\Gamma$
independent of energy.)

In the dynamic case, the simple Breit--Wigner description
must be generalized, and the object to evaluate
is the $S$-Matrix element \cite{JWM94,WJW}. 
Provided interactions in the leads can be neglected, the 
elastic transmission amplitude $t_{\scriptscriptstyle{Q\!D}}(\epsilon)$
can be written as the energy conserving part
of the $S$-Matrix between the left lead and the right lead
\begin{equation}
\lim_{\epsilon' \rightarrow \epsilon} 
{\langle \epsilon', R | {\cal S} | \epsilon, L \rangle} =
\delta(\epsilon' - \epsilon) t_{\scriptscriptstyle{Q\!D}}(\epsilon),
\label{Svst}
\end{equation}
where
\begin{equation}
t_{\scriptscriptstyle{Q\!D}}(\epsilon) = -i \sqrt{\Gamma_L \Gamma_R} 
\langle A(\epsilon,t)\rangle   ,
\label{tvsA}
\end{equation}
with (compare to Sect. \ref{subsec:avecur})
\begin{equation}
\langle{ A(\epsilon,t)}\rangle = \sum_{k=-\infty}^{\infty}
{{J_k^2(V_{\rm ac}/\hbar \omega)}
\over \epsilon - \epsilon_0(V_{\rm s}) - k\hbar\omega + i\Gamma/2}
\;.
\label{Aave}
\end{equation}
\begin{figure}[t]
\psfig{figure=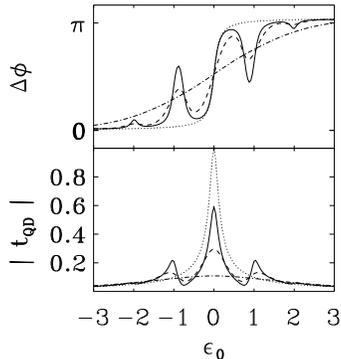,height=1.5in}
\vspace{0.5cm}
\caption{Temperature dependence of the phase shift $\Delta\phi$ (top panel) and
the square of the amplitude (bottom) of 
${t}_{\scriptscriptstyle{Q\!D}}$. 
The level-width is $\Gamma/2=0.1$, in terms of which the other
parameters are $V_{\rm ac}=1.0$, $\omega = 1.0$, and 
$T = 0$ (solid line), $0.1$ (dashed line),  
$0.5$ (dash-dotted line).  For comparison, the $T=0$
time-independent results are shown as dots.  } 
\label{f3}
\end{figure}
The result for the transmission amplitude at finite temperatures is
\begin{equation}
{t}_{\scriptscriptstyle{Q\!D}} =\left(- {\Gamma\over 4\pi T}\right)
\sum_{k=-\infty}^{\infty} J_k^2(V_{\rm ac}/\hbar\omega)
\psi'\left[{1\over 2}-{i\left(\mu-\epsilon_0(V_{\rm s})-
k\hbar\omega+i{\Gamma\over 2}\right)\over 2\pi k_B T}\right]\;,
\label{mainresult}
\end{equation}
where $\psi'$ is the derivative of the digamma function, and
$\mu$ is the chemical potential in the leads.

\begin{figure}[t]
\psfig{figure=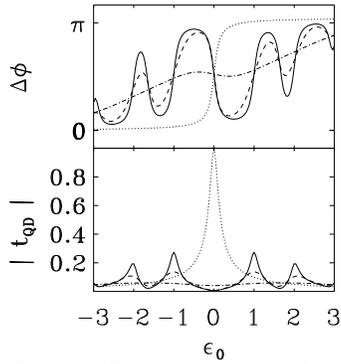,height=1.5in}
\vspace{0.5cm}
\caption{Temperature dependence of the phase shift (top panel) and
the amplitude (bottom) of 
${t}_{\scriptscriptstyle{Q\!D}}$, for
$V_{\rm ac}=2.405$, $\omega=1.0$, $\Gamma/2=0.1$,
$T = 0, 0.1, 0.5$.}
\label{f4}
\end{figure}

Figure \ref{f3} shows the computed magnitude of 
${t}_{\scriptscriptstyle{Q\!D}}$ (bottom)
and its phase (top), as a function of the level energy
$\epsilon_0(V_{\rm s})$.
As compared to the time-independent case (shown as a dotted line),
several features are noteworthy.  The magnitude of
${t}_{\scriptscriptstyle{Q\!D}}$ 
shows photonic side-bands, reminiscent of those seen in transmission
through a microwave modulated quantum dot \cite{Leo}.  
However, there is an important difference from the usual case of photon-assisted
tunneling. The amplitude of the Aharonov--Bohm oscillation is 
sensitive only to the time average of 
the transmission amplitude
$t_{\scriptscriptstyle{Q\!D}}$.
Hence only elastic transmission through the dot contributes,
i.e., the net number of photons absorbed from the ac
field must be zero. The sideband at say $\epsilon = \epsilon_0(V_s) -
\hbar \omega$ corresponds to a process in which an electron
first absorbs a photon to become resonant at energy $\epsilon_0(V_s)$,
and subsequently reemits the photon to return to its original energy.
Perhaps most interesting
are the features appearing in the phase: the phase shift shows a
non-monotonic behavior, with pronounced resonances located at the
energies corresponding to the photonic side-bands.  The strengths
of these phase resonances are strongly dependent on the ac amplitude
$V_{\rm ac}$, and in Fig.\ \ref{f4} we highlight an
interesting consequence of
Eq.(\ref{mainresult}): it is possible to entirely {\it quench} the
main transmission peak (bottom panel), 
or {\it change the sign} of
the slope of the phase at resonance by adjusting 
the ratio $V_{\rm ac}/\hbar\omega$
to coincide with a zero of the Bessel function $J_0$ (top).  
This phenomenon is mathematically
analogous to the recently observed absolute negative conductivity
in THz-irradiated superlattices \cite{Keay}; in our case, however,
it is the {\it phase} rather than the current that displays this behavior.
For a reader interested in further developments along these lines,
we direct attention to two very recent papers \cite{Citrin,Sun3}.

\subsection{Time-dependent Kondo physics}

The Kondo effect is undoubtedly one of the most studied
and best understood problems of many-body physics.
Initially, the theory was developed to explain the increase of
resistivity of a bulk metal with magnetic impurities
at low temperatures \cite{Kondo}.  More recently,
the observation of Kondo effect in several 
experiments on quantum dots\cite{Goldhaber,Cronenwett} has demonstrated
that these systems can serve as an important new tool in the study  of
strongly correlated systems. 
Unlike magnetic impurities in metals, the physical parameters
of the quantum dot can be varied continuously, which allows, for
example, systematic experimental study of the crossover between the Kondo,
the mixed-valence, and the non-Kondo regimes.
These aforementioned experiments
were in good agreement with earlier theoretical predictions
\cite{Glaz,Ng,Hersh,MWL}.  It is therefore natural to ask:
can Kondo physics be probed in time-dependent experiments?
Already before the experiments \cite{Goldhaber,Cronenwett} a
few theoretical papers addressed ac-driven Kondo systems
\cite{Hettler,Schiller,Ngalone,Goldin}, but the last
two years have witnessed a flurry of theoretical activity,
see, e.g., Refs.[\cite{Lopez,Kaminski,Nordlander,Yeyati}].

A full discussion of these papers would merit a review of
its own.  Our present goal is however much more modest. This is a rapidly
developing area and still ridden with some controversies, which
will provide fuel for continued intense research efforts.  Important
for this volume is that
the main technical work-horse has been the nonequilibrium
Green function technique.  It is therefore hoped that the reader
will get a feeling of the present excitement from 
the following brief discussion on
some of the recent achievements and open problems
related to the ac-Kondo problem.

The static Anderson model is solvable with Bethe Ansatz \cite{Tsvelik}
or by quantum Monte Carlo methods \cite{Fye}, but a reliable and
simple method to obtain dynamical properties at low temperatures
in the whole range of interactions $(U/\Gamma)$ is not available.
A number of different methods have been used to attack the
problem.  For example, the non-crossing method (NCA) 
may be used following an exact transformation of the $U=\infty$
Anderson model into a slave-boson Hamiltonian \cite{Bickers}.  
The latter is then
solved self-consistently to second order in the tunneling matrix
elements $V_k$.  The NCA gives reliable results for temperatures
down to $T<T_K$, and its time-dependent formulation has been very
useful in applications to problems in surface physics \cite{Langreth,Shao}.
This is the approach of Ref.\cite{Nordlander}, which studies the temporal
response of a quantum dot which is suddenly shifted into the Kondo
regime by a change of a voltage on a nearby gate.  Thus, the NCA
is applied to a calculation of the nonequilibrium spectral density.
It is suggested that subjecting the quantum dot to a sequence of pulses,
and analyzing the resulting current as a function of the duration
of a pulse, will open a window to the build-up mechanism of the
many-body correlations, responsible for the Kondo effect.

NCA, however, does not give exact results as $T\to 0$, and other
methods are called for.  Finite $U$ perturbation theory works
for the symmetric case, but is known to exhibit anomalies away
from this special case.  An attempt to circumvent these problems is
to construct an effective self-energy, which smoothly interpolates between
known limits $U/\Gamma\to 0$ and $\Gamma/U\to 0$, 
and thus may  eliminate some of the problems
related to perturbative approaches \cite{Yeyati93,Kajueter}.
While a plausible process, interpolation is not rigorous,
and the authors of Ref.\cite{Lopez}, suggest a
nonequilibrium generalization of the Friedel sum rule
to construct a consistency check in the computation
of the nonequilibrium Green functions.
The resulting time-averaged density-of-states exhibits a rich
structure: one may find replicas of the Kondo peak and/or
mean-field peaks.  Also, the conductivity is found to
be strongly affected by the ac-driving, and, interestingly,
that it cannot be described by the Tien--Gordon model \cite{TG}, 
i.e., by single-particle ac-assisted tunneling.
The observability of the satellites of the Kondo peak is,
however, brought into question by Kaminski et al. \cite{Kaminski},
who point out the that the coherence of the many-body
correlations responsible for the Kondo anomaly is fragile
against a {\it spin-flip cotunneling} process, which may
occur already at relatively low frequencies \cite{Nord2000}.
It is clear that experiments exploring the ac-Kondo system
would be extremely welcome! \cite{Kondoexp}

\subsection{Semiconductor-superconductor hybrid systems}
Materials technology has in recent years advanced to a 
stage where high quality hybrid structures can be
fabricated.  By hybrid structures we understand combinations
of normal metals (N) and superconductors (S) together with mesoscopic
structures, such as quantum dots (QD).  Thus, in Fig. \ref{fig:qpc}
either of the contacts, or the quantum dot itself could be
superconducting.  Obviously there are many possible combinations
and here we just discuss one of them, namely a N-QD-S structure,
recently studied by Sun et al.\cite{Sun1,Sun2}.

The superconducting contact allows the possibility of an Andreev
reflection \cite{Andreev}: an electron approaching the 
superconductor with subgap
energy may be transferred into the condensate with a simultaneous
creation of a back-propagating hole.  This process reflects itself
in a number of ways, for example as bound states in a S-N-S system
\cite{Morpurgo}, or the even-odd parity asymmetry and the
Coulomb blockade of the Andreev reflection in S-SQD-S or N-SQD-N
systems \cite{Tuominen1,Eiles,Hekking,Hergenrother}.  Traditionally
rf radiation has been one of the standard probes for studying
superconductivity, and a large  number of experimental results have
been reported on ``simple'' hybird systems.  Here we discuss the
case where a single level (with spin) couples to both radiation
and N and S contacts \cite{Sun2}. (It should be noted that a strongly
interacting QD in this configuration would bring us again to the
Kondo realm \cite{Fazio}.)
The Hamiltonian is thus $H=H_L(t)+H_{\rm cen}(t)+H_R + H_{T,L}
+H_{T,R}$, where
$H_L$, $H_{\rm cen}$ and $H_{T,L}$ are standard but the presence of the
superconductor requires the following modification for the terms referring
to the right-hand side of the system:
\begin{eqnarray}
H_R&=&\sum_{p\sigma}\epsilon_p c^\dagger_{p\sigma} c_{p\sigma}
+
\sum_p\left[\Delta^* c_{p\downarrow} c_{-p,\uparrow}
+\Delta c^\dagger_{-p,\uparrow} c^\dagger_{p,\downarrow}\right]\\
H_{T,R}&=&\sum_{p\sigma}
\left[ V_k e^{i e V_R t} c^\dagger_{p\sigma}d_\sigma
+ V_k^* e^{-i e V_R t} d^\dagger_\sigma c_{p\sigma}\right]\;,
\end{eqnarray}
where the voltage of the superconducting contact $V_R$ appears
in the phase of the tunneling coupling \cite{Cuevas96}.  This
can be viewed as a mean field Hamiltonian leading to the
Bogoliubov--de Gennes equations.

The derivation of the time-dependent current proceeds along
the same lines as in Sect. \ref{sect:current}.  In particular,
the central result Eq.(\ref{jfinal}) is still valid, if one
interprets the central region Green functions as Nambu matrices,
\begin{equation}
{\bf G}^<(t,t')=i\left(\begin{array}{cc}
\langle c^\dagger_\uparrow(t') c_\uparrow(t)\rangle&
\langle c_\downarrow(t') c_\uparrow(t) \rangle\\
\langle c^\dagger_\uparrow(t')c^\dagger_\downarrow (t)\rangle&
\langle c_\downarrow(t') c^\dagger\downarrow(t)
\end{array}\right)\;,
\end{equation}
and analogous definitions hold for the retarded Green functions and
self-energy functionals.  The ensuing calculations are quite complicated,
even in the case of a noninteracting quantum dot \cite{Sun2}, and we focus on
some of the main physical conclusions.  
First of all, one must distinguish on what part of the system the
time-variation is affecting, and whether the photon energy is
bigger or smaller than the energy gap.  As an example, let us consider the
case when $\hbar\omega<\Delta$, and that the external radiation affects
the superconductor (by gauge invariance, this is equivalent to the case
when the normal contact {\it and} the dot are affected, but the superconductor
is not).  The dominating contribution to the current arises from a
photon-assisted Andreev term (PAAT), which has several interesting
properties.  Among these is a possibility of electron pumping 
\cite{SW}: the electrons move from a lower potential to higher
potential by absorbing photons, leading to a negative current
(or, absolute negative conductivity \cite{Keay}).  If, on the
other hand, $\hbar>\Delta$, the normal PAT-processes contribute
substantially.  In general, much interesting structure is
seen as a function of the gate voltage, which can be used
to tune the energy levels in the quantum dot, and the structure
is quite different from the one seen in N--QD--N systems\cite{Leo2}.

\section{Concluding remarks}
In this review we have attempted to give some insight to
recent developments in time-dependent transport in mesoscopic
systems, treated by nonequilibrium Green function methods.
The review is by no means exhaustive: many  important and
interesting problems have not been covered,
such as noise \cite{ChenTing,Cuevas2} or surface
acoustic wave driven transport \cite{Karsten} (and there are many
other topics as well).  The scope of the review has not allowed  an in-depth
treatment of many of the topics, but the author's hope is
that the incomplete treatment has raised the reader's curiosity,
and perhaps thereby attracted new researchers into  this dynamic
field of research.

\section*{Acknowledgments}
It is a pleasure to thank Hartmut Haug and Ned Wingreen for numerous
discussions on nonequilibrium Green functions during various collaborations.
I'm also grateful to my group members M. Brandbyge, J. L{\ae}gsgaard, 
and N. A. Mortensen for careful proof-reading, which has substantially
reduced the number of inconsistensies in the original draft.
The author, of course, bears the full responsibility of the
remaining ones.

\end{document}